\documentclass[prl,
               aps,
               twocolumn,
               nofootinbib, 
               preprintnumbers, 
               superscriptaddress]{revtex4-2}

\usepackage{float}
\usepackage{titlesec}
\usepackage{amsmath,amssymb,slashed,braket}
\usepackage{graphicx}
\usepackage{epstopdf}
\usepackage{float,appendix}
\usepackage{enumerate}
\usepackage{subcaption}
\usepackage{tikz-feynman}
\tikzfeynmanset{compat=1.1.0}
\usepackage{feynmp-auto}
\usepackage[colorlinks=true,
            linkcolor=blue,
            urlcolor=blue,
            citecolor=green,          
            bookmarks=true,
            bookmarksnumbered=true,
            breaklinks=true,
            pdfpagemode=Fullscreen,
            pdfstartview=FitBH]{hyperref}

\usepackage{esint}

\usepackage[normalem]{ulem}
\usepackage{color}

\definecolor{Orange}{cmyk}{0,0.61,0.87,0}
\definecolor{JungleGreen}{cmyk}{0.99,0,0.52,0}
\definecolor{OliveGreen}{cmyk}{0.64,0,0.95,0.40}
\definecolor{Brown}{cmyk}{0,0.81,1,0.60}
\definecolor{RoyalBlue}{cmyk}{0.71,0.53,0,0.12}
\definecolor{Gray}{cmyk}{0,0,0,0.40}
\definecolor{LightPink}{cmyk}{0.0,0.25,0,0}
\definecolor{LLightPink}{cmyk}{0.0,0.10,0,0}
\definecolor{LightBlue}{cmyk}{0.25,0,0,0}
\definecolor{LightGray}{cmyk}{0,0,0,0.2}

\usepackage{xcolor}
\definecolor{gesfpurple}{rgb}{0.47,0.19,0.42}

\definecolor{gesflanse}{rgb}{0.00,0.50,0.50}

\definecolor{gesfblue}{rgb}{0.08,0.42,0.76}

\definecolor{gesfred}{rgb}{1,0,0}

\definecolor{gesfwhite}{rgb}{1,1,1}

\definecolor{gesfblack}{rgb}{0,0,0}

\newcommand{\gapp}[1]{{\hypersetup{linkcolor=red}App.\,\ref{#1}\hypersetup{linkcolor=blue}}}
\newcommand{\geqn}[1]{Eq.\,\hypersetup{linkcolor=blue}(\ref{#1})\hypersetup{linkcolor=blue}}
\newcommand{\gfig}[1]{{\hypersetup{linkcolor=violet}Fig.\,\ref{#1}\hypersetup{linkcolor=blue}}}
\newcommand{\gtab}[1]{{\hypersetup{linkcolor=gesflanse}Tab.\,\ref{#1}\hypersetup{linkcolor=blue}}}

\graphicspath{{figs/}}

\begin{document}

\title{
\Large WIMP Dark Matter from a Natural Discrete Gauge Symmetry in the Standard Model
}

\author{Jie Sheng}
\email{jie.sheng@ipmu.jp}
\affiliation{
Kavli IPMU (WPI), UTIAS, University of Tokyo, Kashiwa, 277-8583, Japan}

\author{Tsutomu T. Yanagida}
\email{tsutomu.tyanagida@gmail.com}
\affiliation{
Kavli IPMU (WPI), UTIAS, University of Tokyo, Kashiwa, 277-8583, Japan}
\affiliation{Tsung-Dao Lee Institute,  Shanghai Jiao Tong University, 201210, China}

\author{Kairui Zhang}
\email{kzhang25@ou.edu}
\affiliation{Homer L. Dodge Department of Physics and Astronomy, University of Oklahoma, 73019, USA}

\begin{abstract}

The internal structure of the Standard Model implies a natural $\mathbb{Z}_4 \times \mathbb{Z}_3$ discrete gauge symmetry. Cancellation of the corresponding Dai--Freed anomalies requires the introduction of three right-handed neutrinos and three additional Majorana fermions $\chi_i$. This gauge symmetry forbids the decay of the lightest fermion $\chi_1$ into Standard Model particles, rendering it automatically stable and providing a dark matter candidate without introducing an \emph{ad hoc}
stabilizing symmetry and domain-wall problem. The mass of $\chi_1$ is generated by the vacuum expectation value of a singlet scalar near the electroweak scale, naturally realizing a weakly interacting massive particle (WIMP) freeze-out scenario. Dark matter annihilation proceeds through scalar mediation, allowing the observed relic abundance to be reproduced while remaining consistent with current direct-detection constraints. 
It naturally realizes the secluded dark matter scenario and can be further tested in the next generation of experiments. 

\end{abstract}

\maketitle 

\section{Introduction} 

Although its presence has been confirmed by a wealth of observations~\cite{Planck:2018vyg}, the origin of dark matter (DM) remains unknown and has become one of the most serious fundamental problems in physics~\cite{Cirelli:2024ssz}. A wide variety of DM models have been proposed~\cite{Bertone:2004pz,Bergstrom:2009ib}; among them, the most mainstream and compelling candidates are often strongly motivated by underlying theoretical structures or by unresolved puzzles within the Standard Model (SM). Two of the best-known examples are the lightest supersymmetric partner~\cite{Arcadi:2017kky,Baer:2025srs,Baer:2025zqt,Zhang:2026eoc} of an SM particle—commonly referred to as a WIMP DM—and the axion DM proposed to solve the strong CP problem~\cite{Peccei:1977ur,Peccei:1977hh,Wilczek:1977pj,Weinberg:1977ma,Baer:2026wre,Baer:2026zra}.

In this paper, we propose a new DM candidate that is closely tied to a deep structural feature of the SM. Recall that the SM admits two types of instanton solutions: QCD instantons and electroweak instantons. Remarkably, both solutions possess 12 fermion zero modes~\cite{Nomura:2000yk}, implying that a $\mathbb Z_{12}=\mathbb Z_4\times \mathbb Z_3$ symmetry is embedded in the SM. It is therefore well motivated to interpret this $\mathbb Z_{12}$ as a discrete gauge symmetry, since it is free from anomalies associated with the SM gauge interactions. 

However, recent studies have shown that discrete gauge symmetries are also subject to non-perturbative consistency conditions in the form of \textit{Dai–Freed anomalies}~\cite{Dai:1994kq,Yonekura:2016wuc}. Interestingly, all such anomalies are canceled if the SM is extended by three right-handed neutrinos (RHNs) $\bar N_i$ and three additional Majorana fermions $\chi_i$ with $i=1,2,3$~\cite{Sheng:2025sou}. As will be shown, the lightest $\chi_1$ is automatically stable under such symmetry, and it thus plays the role of DM. Moreover, our framework predicts that the DM candidate $\chi_1$ can be thermally produced as a WIMP and can be probed in future Xenon-based direct-detection experiments. This provides a theoretically well-motivated realization of the WIMP paradigm without supersymmetry (SUSY).

\section{WIMP DM from Anomaly-Free Discrete Gauge Symmetries}

Gauge symmetry is the cornerstone of our understanding of elementary particles and their interactions, and the SM of particle physics is built upon gauge symmetries, $SU(3)_c\times SU(2)_L \times U(1)_Y$. To address various beyond-the-Standard-Model (BSM) problems, the most widely pursued strategy is to search for new symmetries that are closely related to those of the SM. 

The discrete gauge symmetry $\mathbb Z_4$ is a successful example motivated by the internal structure of the SM. 
Since in a single fermion generation there are four Weyl fermions (the left- and right-handed up and down quarks) contributing to the QCD anomaly, and four $SU(2)_L$ doublets (the three colored quark doublets and the lepton doublet) contributing to the electroweak anomaly, the SM naturally admits an anomaly-free $\mathbb Z_4$ symmetry when all chiral fermions are assigned the corresponding charge $+1$. However, this anomaly cancellation is spoiled once non-pertuebative effects are taken into account, 
which is known as the \textit{Dai–Freed anomaly}~\cite{Dai:1994kq,Yonekura:2016wuc}. Consistency then requires extending the SM by three right-handed neutrinos $\bar N_i$~\cite{Garcia-Etxebarria:2018ajm,Kawasaki:2023mjm}, which constitute a minimal
and well-motivated extension of the theory. These fields naturally account for the observed small neutrino masses via the seesaw mechanism \footnote{The term \textit{seesaw mechanism} was first coined by one of the authors, T. T. Y., at the INS symposium in Tokyo 1981~\cite{INS1981}, where the dimension 5 operator for neutrino masses is also pointed out.}~\cite{Minkowski:1977sc,Yanagida:1979as,Yanagida:1979gs,Gell-Mann:1979vob}, and generate the present baryon asymmetry through leptogenesis~\cite{Fukugita:1986hr}.

Motivated by the above success of the $\mathbb Z_4$ symmetry, it is natural to ask what other embedded symmetries may exist in the SM. A simple observation is that there are three generations of fermions, so a discrete gauge symmetry $\mathbb Z_3$ is also anomaly-free with respect to all SM gauge interactions if all chiral fermions are assigned a $\mathbb Z_3$ charge of $+1$. 
The origin of both $\mathbb Z_4$ and $\mathbb Z_3$ gauge symmetry in SM and the charge assignments are discussed in \gapp{appendix-discrete}
A discrete $\mathbb Z_3$ gauge symmetry also suffers from the Dai–Freed anomaly~\cite{Hsieh:2018ifc}, which can again be canceled by introducing three chiral fermions, $\chi_i$, whose $\mathbb Z_3$ charge is $-1$ and the lightest of them serves as DM. This prediction is highly nontrivial and phenomenologically important~\cite{Sheng:2025sou,Georis:2025kzv}. In this work, we show how the discrete $\mathbb Z_3$ symmetry, together with the new fields $\chi$, answers the question of \textit{who orders the WIMP DM?}.

For notational convenience, we organize the SM matter fields into $SU(5)$ representations, 
\begin{subequations}
\begin{align}
    T ({\bf 10}) & \equiv \{q, \bar{u}, \bar{e} \} , \\
    \bar{F} ({\bf 5}^*) & \equiv \{\bar{d}, \ell\},  \\
   H({\bf 5}^*) &\equiv \{H_1, H_2\}.
\end{align}
\end{subequations}
purely as a bookkeeping device. No grand unification of the SM gauge groups is assumed, nor are any GUT relations imposed on the gauge or Yukawa couplings. All fermions are written as left-handed Weyl spinors. The $\mathbb Z_4 \times \mathbb Z_3$ charge assignments for all fermions are summarized in the following \gtab{fermion}\footnote{The proof of such a charge assignment is shown in Ref.~\cite{Hsieh:2018ifc}. The anomaly-free condition of $\mathbb Z_n$ discrete gauge symmetry is that both $(n^2 + 3n +2) S_3 /(6n)$ and $2 S_1/n$ are integers. Since we write all the fermion fields as the left-handed Weyl fields, here the $S_1$ and $S_3$ are defined as $S_1 = \sum_i s_i$ and $S_3 = \sum_i s^3_i$ where $s_i$ is the corresponding charge. In SM, we have $15$ Weyl fields with $3$ generations. $\mathbb Z_4$ symmetry requires three generations of the right-handed neutrinos~\cite{Kawasaki:2023mjm}. One can check that if the $\chi$ field has $3$ generations and charge $-1$, or equivalently $+2$ under $\mathbb Z_3$, could cancel the anomaly.
}. 

\begin{table}[H]
    \centering
    \large
    \begin{tabular}{p{2cm} p{1cm} p{1cm} p{1cm} p{1cm}}
    \hline
    Fermions & $T$ & $\bar F$ & $\bar N$ & $\chi$ \\
    \hline
    \hline
    \qquad $\mathbb{Z}_4$ & 1 & 1 & 1 & 0 \\
    \hline
    \qquad$\mathbb{Z}_3$ & 1 & 1 & 1 & -1 \\
    \hline
    \end{tabular}
    \caption{The charges of fermions under $\mathbb Z_4$ and $\mathbb Z_3$ symmetries.}
        \label{fermion}
\end{table}

In order to reproduce the SM fermion mass terms under the discrete gauge symmetry $\mathbb Z_3$, a single Higgs field is not enough and two Higgs doublets $H_1$ and $H_2$ are naturally required to write down the following Yukawa interactions~\cite{Sheng:2025sou},
\begin{equation} 
    \mathcal{L}_{\text{Y}}^{\text{SM}} \supset \frac{1}{2}T TH_1^\dagger + T \bar{F} H_2 + 
    \bar N\bar FH_1^\dagger  + \text{h.c.}.
\end{equation}
It can be expanded as the standard mass terms for quark and lepton fields as,
\begin{equation}
    \mathcal{L}_{\text{Y}}^{\text{SM}} \supset
     - f_u \bar u  H_1^\dagger {q}
        - f_e \bar{e} H_2 \ell
        - f_d \bar d H_2 q
        - f_N \bar N H_1^\dagger \ell
        + \text{h.c.}.
\label{SM_yukawa}
\end{equation}
The discrete gauge charges assignments of $H_1$ and $H_2$ are show in~\gtab{higgs}. An accidental $U(1)$ symmetry can naturally appear within the two-Higgs-doublet extension of the SM~\cite{Peccei:1977ur,Peccei:1977hh}. It can be identified as the so called Peccei-Quinn symmetry~\cite{Peccei:1977ur,Peccei:1977hh}. However, this global symmetry would be explicitly broken by the following introduced scalar fields $\Phi$ and $\varphi$.
\begin{table}[h] 
    \centering 
    \large 
    \begin{tabular}{p{1.5cm} p{1cm} p{1cm} p{1cm} p{1cm}} 
        \hline 
        Scalars & $H_1$ & $H_2$ & $\Phi$ & $\varphi$ \\ 
        \hline 
        \hline 
        $\quad \mathbb{Z}_4$ & -2 & -2 & -2 & 0  \\ 
        \hline 
        $\quad \mathbb{Z}_3$ & 2 & 1 & -2 & -1 \\ 
        \hline 
    \end{tabular} 
\caption{The charges of all scalar fields under $\mathbb Z_4$ and $\mathbb Z_3$ symmetries.}\label{higgs} 
\end{table}

In order to generate mass terms for the RHNs $\bar N_i$ and the Majorana fermions $\chi_i$ without Planck suppression, two scalar fields are needed: $\Phi$ for $\bar N_i$ and $\varphi$ for $\chi_i$. Adopting the principle of minimality, the allowed Yukawa interactions are
\begin{equation} 
    \mathcal{L}_{\text{Y}}^{N,\chi} \supset -\frac{1}{2} y_{N_i} \Phi \bar N_i \bar N_i - \frac{1}{2} y_{\chi_i} \varphi \chi_i \chi_i - y_{ij} \frac{\Phi^2}{M_{\text{pl}}}\chi_i \chi_j + \text{h.c.},
\label{eq:massterms}
\end{equation}
where all higher-dimensional operators are suppressed by the reduced Planck scale $M_{\text{pl}} \simeq 2.4 \times 10^{18}\,\text{GeV}$, which is the only cutoff scale. Once the scalar fields $\Phi$ and $\varphi$ acquire vevs, they generate Majorana
masses for $N_i$ and $\chi_i$, leading to $m_{N_i} \equiv y_{N_i} \braket{\Phi} \equiv y_{N_i} v_\Phi$ and $m_{\chi_i} \equiv y_{\chi_i} \braket{\varphi} \equiv y_{\chi_i} v_\varphi$, respectively. 

The decay of RHNs in the early Universe generates a lepton asymmetry, which is subsequently converted into a baryon asymmetry by sphaleron processes~\cite{Fukugita:1986hr}. The model therefore also accounts for the observed baryon asymmetry via leptogenesis. In the standard hierarchical thermal leptogenesis scenario, the Davidson--Ibarra bound~\cite{Davidson:2002qv,Davidson:2008bu} requires the lightest RHN to be heavier than $\sim 10^9~\text{GeV}$, which in turn, puts a lower bound on $v_\Phi \gtrsim 10^{9}~\text{GeV}$, and thus, on reheating temperature\footnote{The two Higgs doublets structure in~\geqn{SM_yukawa} modifies the Davidson-Ibarra bound slightly to be $m_{N} \gtrsim 10^{9} \times \left(v_1/246~\text{GeV}\right)^2$ GeV~\cite{Huang:2024azp}.}.

After spontaneous symmetry breaking due to the vev of scalar field $\Phi$, the discrete gauge structure leaves a residual $\mathbb{Z}_2$ symmetry. Since $\Phi$ has $\mathbb Z_4$ charge $-2$, its vev preserves a non-trivial $\mathbb Z_2$ subgroup of $\mathbb Z_4$. Owing to this residual $\mathbb{Z}_2$ symmetry, the lightest state $\chi_1$ is stable. The field $\chi$ is even under the $\mathbb{Z}_2$ symmetry, while all SM fermions are odd. Therefore, any operator containing a single $\chi$ and an odd number of SM fermions is odd under this residual $\mathbb{Z}_2$ and is forbidden. Since $\chi_1$ is the lightest fermion that is even under the $\mathbb Z_2$, it cannot decay and is automatically stable. 
The particle $\chi_1$ therefore naturally serves as a DM candidate, without the need to impose an \emph{ad hoc} stabilizing symmetry.

Since $\chi$ acquires its mass from the vev $v_\varphi$ and couples directly only to the scalar $\varphi$, avoiding overclosure of the Universe by $\chi$ requires $v_\varphi$ to lie near the electroweak scale. As will be shown below, this is required to have sufficiently efficient annihilation of $\chi$ into SM particles through $\varphi$-$h$ mixing prior to its freeze-out. The model thus naturally realizes a WIMP DM scenario, which constitutes a central prediction of our model.

The gauge charge assignments also allow the following interaction terms among
$\varphi$ and $\Phi$, and the Higgs doublets $H_{1,2}$:
\begin{equation}\label{eq:scalar_int}
\begin{split}
    -&\mathcal{L}^{\text{scalar}}_\text{int} 
        \supset
            [\mu_\varphi \varphi (H_2^\dagger H_1 ) + g_{\varphi} \varphi^2 (H_1^\dagger H_2) \\
                &\qquad+ g_\Phi \Phi^2 (H_2^\dagger H_1 ) + \text{h.c.}] \\
                &\qquad\quad + \lambda_{\varphi i}|\varphi|^2 (H_i^\dagger H_i) + \lambda_{\Phi i}|\Phi|^2 (H_i^\dagger H_i) \\
                &\qquad\quad + \mu'\Phi^2 \varphi^* + \cdots,
\end{split}
\end{equation}
where $i=1,2$, and the ellipsis denotes the other scalar self-interaction terms
$V(H_1,H_2)$ and $V(\Phi,\varphi)$. There also exist other interaction terms between $\varphi$ and $\Phi$, but they are irrelevant for the phenomenology of the dark matter particle $\chi$ and are therefore omitted.

Maintaining the scalar $\varphi$ at the weak scale requires its quadratic terms, including both the bare mass and contributions induced by $\Phi^2$, to be parametrically small compared to the Planck scale. However, a weak-scale $\varphi$ (and hence a weak-scale $\chi$) is singled out by the requirement of successful thermal DM
production and non-overclosure of the universe as shown later. From this perspective, the smallness of the $\Phi$--$\varphi$ couplings is viewed as environmentally selected, in line with Weinberg's anthropic arguments~\cite{Weinberg:1987dv}. 

Interestingly, since both discrete symmetries are already spontaneously broken by $\Phi$ at a high scale, stable domain walls, if formed at high energies, are washed out by inflation. Moreover, the term $\mu'\langle\Phi\rangle^2 \varphi^*$ in~\geqn{eq:scalar_int} acts as an effective tadpole for $\varphi$ at low energies, explicitly lifting the would-be degenerate vacua and preventing the formation of stable domain walls at low scales as well. To ensure that the walls collapse before they dominate the energy density, the tadpole should at least satisfy $\mu'\langle\Phi\rangle^2 \gtrsim v_{\text{EW}}^4/M_{\text{Pl}}$~\cite{Stewart:1996ai},
which is consistent with the fact that any mass scales for $\varphi$ are at the weak scale. The model therefore preserves the stability of $\chi_1$ without introducing a domain-wall problem.

Finally, the large value of $v_\Phi$ suggests that all additional Higgs states associated with $\Phi$ are very heavy, except for the observed 125 GeV Higgs boson. Consequently, the low-energy scalar sector effectively reduces to a SM-like
single-Higgs + a SM singlet scalar $\varphi$ theory at the weak scale.
Our framework thus provides a theoretically consistent and motivated model for secluded-portal WIMP DM~\cite{Pospelov:2007mp,Ellwanger:2009dp,Arcadi:2016qoz,Arcadi:2017kky,Meng:2024lmi,DiMauro:2025jsb}, in which the dark-sector mediator couples to SM fermions only through additional mixing, such as mixing with the Higgs boson~\cite{Kim:2008pp,Lopez-Honorez:2012tov,Esch:2013rta,Bell:2016ekl,Bell:2017rgi}. As shown below, the DM can be thermally produced with the correct relic abundance while evading current direct-detection constraints.

\section{Light Scalars Mixing}
The interactions between the light singlet scalar $\varphi$ and the Higgs doublets $H_1$ and $H_2$ induce mixing between the CP-even component of $\varphi$ and the SM-like Higgs boson~\cite{Kim:2008pp,Lopez-Honorez:2012tov,Esch:2013rta,Falkowski:2015iwa,Bell:2016ekl,Bell:2017rgi}. In the decoupling limit of the two-Higgs-doublet sector, the light CP-even Higgs
state is given by
\begin{equation}
    h_0 \equiv \cos\beta H_1^0 + \sin\beta H_2^0,
\end{equation}
where $H_i^0$ denotes the neutral component of the Higgs doublet $H_i$, $\tan\beta \equiv v_2/v_1$, and $v_\text{EW}^2 = v_1^2 + v_2^2 = (246~\text{GeV})^2$. In the absence of mixing with $\varphi$, the state $h_0$ coincides with the observed Higgs boson $h$ of mass $125~\text{GeV}$.

The scalar potential for $\varphi$ is parametrized as
\begin{equation}\label{eq:varphi_potential}
    V(\varphi) = -m^2_\varphi|\varphi|^2 + \frac{\kappa}{3!}(\varphi^3 + \varphi^{*3}) + \frac{\lambda}{4}|\varphi|^4.
\end{equation}
Expanding $\varphi$ around its vev yields
\begin{equation}
    \varphi = \frac{1}{\sqrt{2}}(v_\varphi + s_0 + ia).
\end{equation}
Since the potential in~\geqn{eq:varphi_potential} does not possess an accidental global $U(1)$ symmetry, the would-be Goldstone boson $a$ acquires a mass. Its mass is parametrically set by the trilinear coupling, $m_a^2 \sim -\kappa v_\varphi$, and can be taken sufficiently large to decouple from the low-energy spectrum. As an independent parameter of the scalar potential, $\kappa$ is not required to be of the same order as $v_\varphi$.

In the decoupling limit, all non-SM Higgs states ($H_0$, $A$, $H^\pm$) as well as scalar excitations associated with the field $\Phi$ are assumed to be heavy and can be safely integrated out. The low-energy CP-even scalar sector is therefore described by the two fields $(h_0, s_0)$. From~\geqn{eq:scalar_int}, the corresponding mass-squared matrix takes the form
\begin{equation}
    M^2 = 
        \begin{pmatrix}
            m^2_{hh} & m^2_{hs} \\
            m^2_{hs} & m^2_{ss}
        \end{pmatrix},
\end{equation}
where 
\begin{gather}
    m^2_{hh} \simeq \frac{1}{2}\left[m^2_{h0} + (\lambda_{\varphi 1}c^2_\beta + \lambda_{\varphi 2}s^2_\beta)v_\varphi^2 + g_{\varphi}v_\varphi^2s_{2\beta} \right], \\
    m^2_{ss} \simeq \frac{1}{2}\left[m^2_{s0} +(\lambda_{\varphi 1}c^2_\beta + \lambda_{\varphi 2}s^2_\beta)v_\text{EW}^2 + g_{\varphi}v_\text{EW}^2s_{2\beta} \right], \\
    m^2_{hs} \simeq v_\text{EW}v_\varphi(\lambda_{\varphi 1}c^2_\beta + \lambda_{\varphi 2}s^2_\beta + g_{\varphi}s_{2\beta}) + \frac{\mu_\varphi v_\text{EW}}{\sqrt{2}}s_{2\beta}.
\end{gather}
Here we define $s_\beta \equiv \sin\beta$, $c_\beta \equiv \cos\beta$, and
$m_{s0}^2 \equiv \left.\partial^2 V(\varphi)/\partial s_0^2\right|_{v_\varphi}$,
while $m_{h0}^2$ denotes the Higgs mass parameter prior to mixing.

Diagonalizing $M^2$ by a rotation with mixing angle $\theta$~\cite{Falkowski:2015iwa,Arcadi:2017kky},
\begin{equation}
    \begin{pmatrix}
        h \\ s
    \end{pmatrix}
    =
    \begin{pmatrix}
        \cos\theta & \sin\theta \\
        -\sin\theta & \cos\theta
    \end{pmatrix}
    \begin{pmatrix}
        h_0 \\ s_0
    \end{pmatrix},
\end{equation}
yields
\begin{equation}
    \tan 2\theta = \frac{2 m_{hs}^2}{m_{ss}^2 - m_{hh}^2},
\end{equation}
with eigenvalues
\begin{equation}
    m^2_{1, 2} = \frac{1}{2}\left(m^2_{hh} + m_{ss}^2 \mp \sqrt{(m^2_{ss} - m^2_{hh})^2 + 4m^4_{hs}}\right).
\end{equation}
We identify one of the eigenstate with the observed Higgs boson,
\begin{equation}
    m_{h, obs} \equiv m_1 = 125\, \text{GeV},\quad m_s \equiv m_2.
\end{equation}
The couplings of the physical scalars $h$ and $s$ to the dark matter particle
$\chi$ are rescaled by the mixing angle,
\begin{equation}\label{eq:g_scalar_chi}
    g_{h\chi\chi} = \frac{y_{\chi_1}}{2\sqrt{2}}s_\theta = \frac{m_\chi}{v_\varphi}s_\theta,\quad g_{s\chi\chi} = \frac{y_{\chi_1}}{2\sqrt{2}}c_\theta = \frac{m_\chi}{v_\varphi}c_\theta,
\end{equation}
where we used the shorthands $\cos\theta \equiv c_\theta$ and $\sin\theta \equiv s_\theta$. Similarly, the couplings to SM particles $X$ are given by
\begin{equation}\label{eq:g_scalar_X}
    g_{hXX} = c_\theta g_{hXX}^\text{SM},\quad g_{sXX} = -s_\theta g_{hXX}^\text{SM},
\end{equation}
where the SM Higgs couplings are
\begin{equation}\label{eq:g_SM}
    g^\text{SM}_{hff} \equiv \frac{m_f}{v_\text{EW}}, \quad g^\text{SM}_{hVV} \equiv \frac{2m_V^2}{v_\text{EW}}.
\end{equation}
In the following analysis, we focus on the phenomenologically relevant regime
$\sin\theta \ll 1$, consistent with collider constraints such as limits from
gauge-boson-fusion production of $s$ followed by invisible decays~\cite{ATLAS:2015ciy}. Also, we only consider $m_s > m_h /2 \simeq 62.5$ GeV to avoid the direct Higgs invisible decay bound via $h\to ss$~\cite{ATLAS:2015ciy}.

\section{Thermal Dark Matter Genesis}
In this model, the WIMP-like DM particle $\chi_1$ is naturally thermally produced in the early Universe and can account for the entire observed relic abundance, $\Omega_{\rm DM} \sim 0.1$. Thermal WIMP DM is phenomenologically well motivated and is commonly discussed in the context of the so-called \emph{WIMP miracle}.

\begin{figure}[htb!]
    \centering
    \vspace{0.1cm}
    \begin{subfigure}{0.23\textwidth}
        \centering
        \includegraphics[width=\textwidth,trim=10 0 10 0,clip]{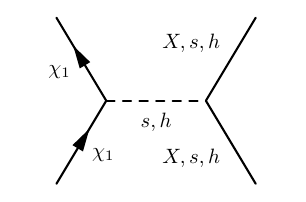}
        \caption{}
        \label{fig:chi-11XX-s}
    \end{subfigure}
    \begin{subfigure}{0.23\textwidth}
        \centering
        \includegraphics[width=\textwidth,trim=10 0 10 0,clip]{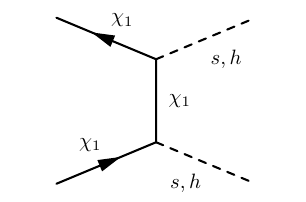}
        \caption{}
        \label{fig:chi-11XX-t}
    \end{subfigure}
    
    \caption{Number changing diagrams involving $\chi_1$ that are relevant to DM genesis, where $X$ denotes the possible SM fermion and vector boson final states.
     }
    \label{fig:chi1_diagrams}
\end{figure}
\begin{figure}[htb!]
    \newcommand{\comma}{,}
    \begin{subfigure}{0.23\textwidth}
        \centering
        \includegraphics[width=\textwidth,trim=10 0 10 0,clip]{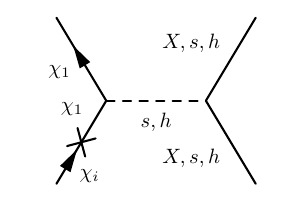}

        \caption{}
    \end{subfigure}
    \begin{subfigure}{0.23\textwidth}
        \centering
        \includegraphics[width=\textwidth,trim=10 0 10 0,clip]{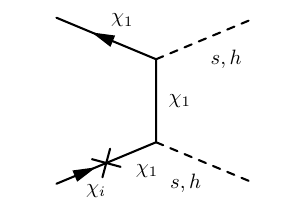}

        \caption{}
    \end{subfigure}

    \begin{subfigure}{0.23\textwidth}
        \centering
        \includegraphics[width=\textwidth,trim=10 0 10 0,clip]{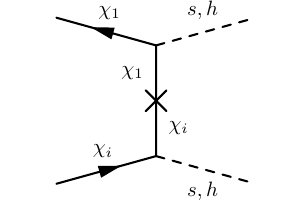}

        \caption{}
    \end{subfigure}
    
    \caption{Number changing diagrams involving coannihilating $\chi_1$ and $\chi_i$. The cross denotes the effective intra-generation mixing among $\chi_i\chi_1$ induced by $\Phi$.
     }
    \label{fig:chi1_coa_diagrams}
\end{figure}

Through the number-changing processes shown in \gfig{fig:chi1_diagrams} and \gfig{fig:chi1_coa_diagrams}, $\chi_1$ efficiently thermalizes with the SM plasma in the early Universe, unless some couplings are tuned to be much smaller than $\mathcal{O}(1)$. Note our scenario is also similar to that in secluded DM scenarios~\cite{Arcadi:2017kky}. For a thermal relic to survive, these interactions must fall out of equilibrium at the freeze-out temperature $T_{\rm f.o.}$, satisfying $\Gamma_{\chi_1}(T_{\rm f.o.}) \lesssim H(T_{\rm f.o.})$. In the WIMP framework, the relic abundance can be well approximated by
\begin{equation}\label{eq:wimp_miracle}
    \Omega_\text{DM} 
        \simeq 0.1\left( \frac{x_\text{f.o.}}{20} \right) \left( \frac{10^{-8}~\text{GeV}^{-2}}{\langle\sigma v\rangle} \right),
\end{equation}
where $x \equiv m_{\chi_1}/T$ and $\braket{\sigma v}$ denotes the
thermally averaged annihilation cross section.

The freeze-out parameter $x_{\rm f.o.}$ can be obtained iteratively as~\cite{Kolb:1990vq}
\begin{equation}
    x_\text{f.o.} \equiv \ln \frac{0.038 g_{\chi_1}M_\text{pl}m_{\chi_1} \braket{\sigma v}}{g^{1/2}_*x_\text{f.o.}^{1/2}},
\end{equation}
where $g_{\chi_1} = 2$ for a Majorana fermion and $g_*$ denotes the number of relativistic degrees of freedom at decoupling. For most of the parameter space relevant to thermal freeze-out, $x_{\rm f.o.} \sim 20$--$30$.

We focus primarily on the DM mass range $m_{\chi_1} \sim \mathcal{O}(10^2)$--$\mathcal{O}(10^5)\,$GeV. The lower part of this range is already strongly constrained by direct detection experiments, as shown in \gfig{DD}\footnote{Below $10\,$GeV, one gradually departs from the traditional WIMP mass range and becomes subject to CMB constraints~\cite{Kawasaki:2021etm}. Nevertheless, we will also briefly discuss the parameter space below $100\,$GeV throughout this section.}, while the upper bound is set by the unitarity limit on annihilation cross sections~\cite{PhysRevLett.64.615,vonHarling:2014kha,Smirnov:2019ngs}, unless exotic annihilation channels are introduced~\cite{Harigaya:2016nlg,Kramer:2020sbb,Frumkin:2022ror,Qiu:2023bbp}.

The relevant $\chi_1\chi_1$ annihilation cross sections contributing to \gfig{fig:chi1_diagrams} are listed below, keeping terms up to $\mathcal{O}(\sin^2\theta)$. Since $\chi_1$ is non-relativistic at freeze-out, we perform a velocity
expansion $\langle\sigma v\rangle \simeq a + b v^2$, or equivalently $\langle\sigma v\rangle \simeq a + 3b/(2x)$ \cite{Gondolo:1990dk}. This approximation is valid when these processes are away from $s$-channel resonances and kinematic thresholds.
\begin{equation}
\begin{split}
    \braket{\sigma v}_{ss}&
        = \frac{3\sqrt{1 - r_s}}{128x\pi} \Bigg[\frac{c_\theta^8\tilde{\mu}_{sss}^2}{2m_{\chi_1}^2v^2_\varphi(4 - r_{s})^2} \\
                &\quad + \frac{8m^2_{\chi_1} c_\theta^4}{3v_\varphi^4}\frac{(1 - r_s)^2}{(2 - r_s)^4} + \frac{4c_\theta^6 \tilde{\mu}_{sss}}{3v^3_\varphi} \frac{(1-r_s)}{(4 - r_{s})(2 - r_s)^2} \Bigg],
\label{eq:xsec_ss}
\end{split}
\end{equation}
\begin{equation}
\begin{split}
    \braket{\sigma v}_{hh}
        = \frac{3\sqrt{1 - r_h}}{256x\pi} \frac{c_\theta^6s^2_\theta\tilde{\mu}_{hhh}^2}{m_{\chi_1}^2v^2_\varphi(4 - r_h)^2},
\label{eq:xsec_hh}
\end{split}
\end{equation}
\begin{equation}
\begin{split}
    \braket{\sigma v}_{hs}
        &= \frac{3\sqrt{\lambda(4, \sqrt{r_s}, \sqrt{r_h})}}{256\pi x } \\ &\quad\times\Bigg[\frac{c_\theta^6 s_\theta^2\tilde{\mu}^2_{ssh}}{2m_{\chi_1}^2v^2_\varphi(4 - r_{s})^2} 
            + \frac{m^2_{\chi_1} c_\theta^2 s_\theta^2}{6v_\varphi^4}\frac{\lambda^2(4, \sqrt{r_s}, \sqrt{r_h})}{(4 - r_s - r_h)^4} \\
                &\qquad\qquad + \frac{c_\theta^4s_\theta^2 \tilde{\mu}_{ssh}}{3v^3_\varphi} \frac{\lambda(4, \sqrt{r_s}, \sqrt{r_h})}{(4 - r_{s})(4 - r_s - r_h)^2}\Bigg],
\label{eq:xsec_hs}
\end{split}
\end{equation}
where $\lambda(x,y,z) \equiv x^2 + (y-z)^2 - 2x(y+z)$ is the Källén function, and
$r_{s,h,X} \equiv m_{s,h,X}^2/m_{\chi_1}^2$.
In the small-mixing limit, the relevant scalar trilinear couplings can be
approximated at leading order in $\theta$ as
\begin{equation}
    \frac{\mu_{sss}}{3!} \approx \frac{1}{3!}\left(\frac{\kappa}{\sqrt{2}} + 3\frac{m_s^2}{v_\varphi}\right)c^3_\theta \equiv \frac{\tilde{\mu}_{sss}}{3!} c^3_\theta,
\end{equation}
\begin{equation}
    \frac{\mu_{ssh}}{2!}
        \approx \frac{3}{3!}\left( \frac{\kappa}{\sqrt{2}} + 3\frac{m_s^2}{v_\varphi} \right) c^2_\theta s_\theta 
        \equiv \frac{\tilde{\mu}_{ssh}}{2!}c^2_\theta s_\theta,
\end{equation}
and
\begin{equation}
    \frac{\mu_{hhh}}{3!} 
        \approx \frac{1}{3!}\left(3\frac{m_h^2}{v_\text{EW}}\right) c^3_\theta
        \equiv \frac{\tilde{\mu}_{hhh}}{3!}c^3_\theta.
\end{equation}
Here the $\tilde{\mu}$ are defined by factoring out the explicit
$\theta$ dependence. The Majorana property of $\chi_1$ gives the $p$-wave annihilation cross section, as expected.

In addition to the scalar final states, $\chi_1$ can annihilate into SM fermions $f$ and massive vector bosons $V$ through $s$-channel exchange of the mixed scalars $h$ and $s$ as shown in \gfig{fig:chi1_diagrams}. The corresponding thermally averaged cross sections are
\begin{equation}
    \braket{\sigma v}_{ff} = \sum_f\frac{3N_c(1 - r_f)^{3/2}}{16x\pi}\frac{m_f^2 s_\theta^2 c_\theta^2}{v_{\varphi}^2 v_\text{EW}^2} \left( \frac{1}{4 - r_s} - \frac{1}{4 - r_h} \right)^2,
\label{eq:xsec_ff}
\end{equation}
where $N_c$ is the color factor, and
\begin{equation}
\begin{split}
    \braket{\sigma v}_{VV}
        &= \frac{3\sqrt{1 - r_V}}{32x\pi S}\frac{s_\theta^2 c_\theta^2(4m_\chi^4 - 4m_\chi^2m_V^2 + 3m_V^4)}{m_\chi^2 v^2_\varphi v^2_h} \\
            &\qquad \times\left( \frac{1}{4 - r_s} - \frac{1}{4 - r_h} \right)^2,
\label{eq:xsec_VV}
\end{split}
\end{equation}
with $S=2$ for $V=Z$ and $S=1$ for $V=W$, accounting for identical particles in
the final state.

\begin{figure*}[htb!]
    \centering
    \begin{subfigure}{0.49\textwidth}
        \centering
        \includegraphics[width=\textwidth]{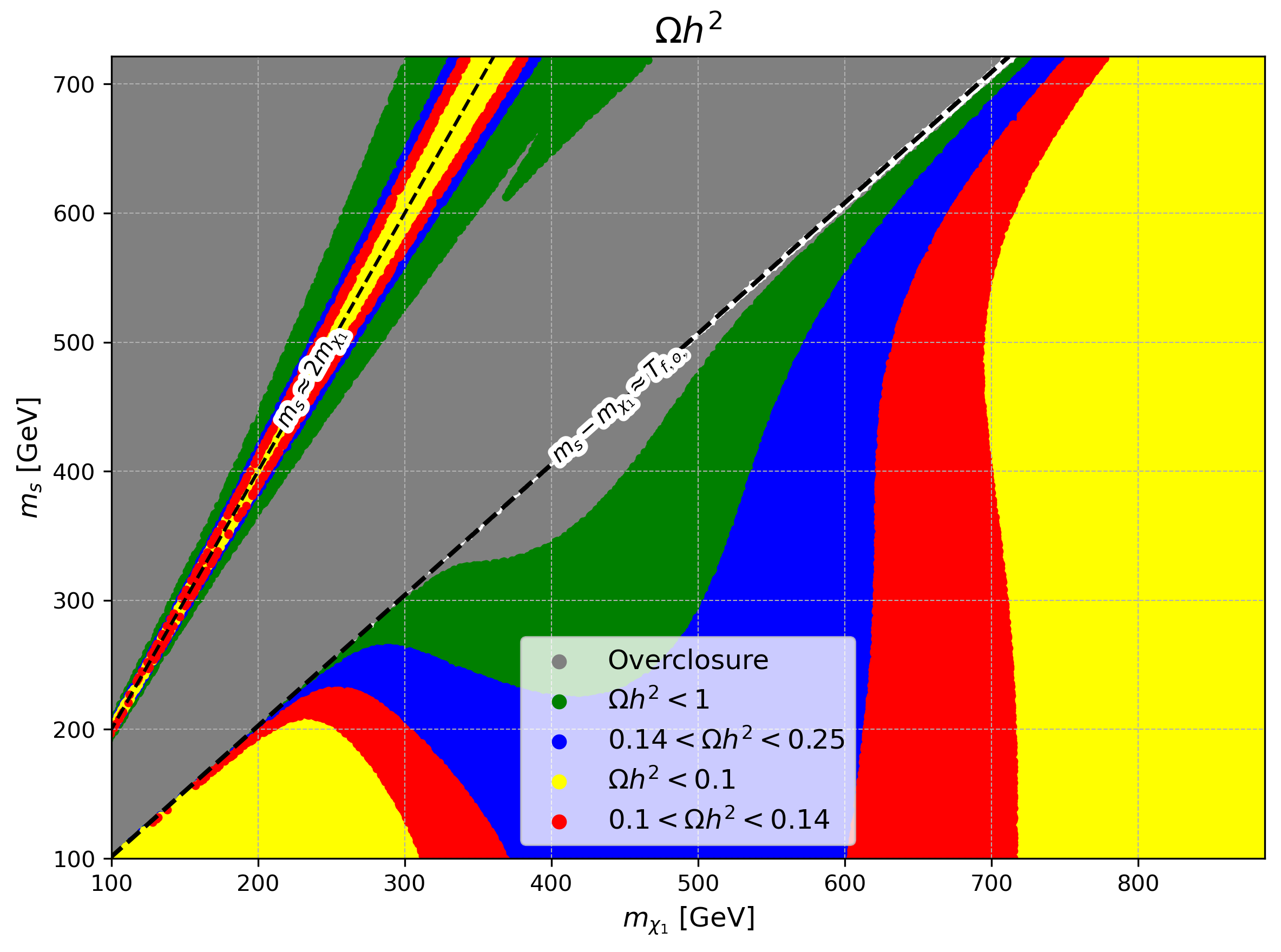}
        \caption{$v_\varphi = 0.25$ TeV, $\kappa = -3$ TeV, $\sin\theta = 0.05$}
    \end{subfigure}
    \begin{subfigure}{0.49\textwidth}
        \centering
        \includegraphics[width=\textwidth]{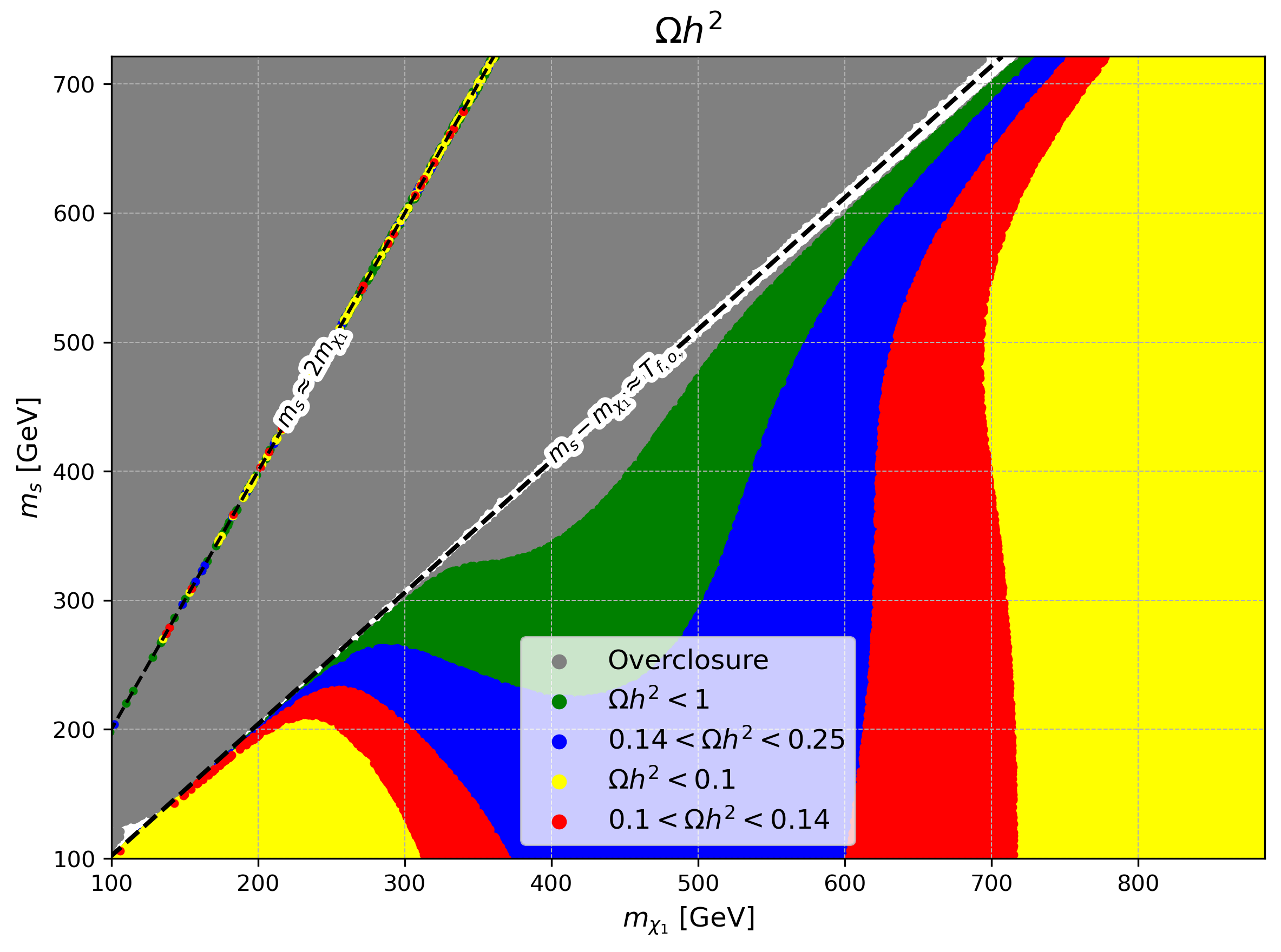}
        \caption{$v_\varphi = 0.25$ TeV, $\kappa = -3$ TeV, $\sin\theta = 10^{-3}$}\label{fig:relic_b}
    \end{subfigure}
    
    \begin{subfigure}{0.49\textwidth}
        \centering
        \includegraphics[width=\textwidth]{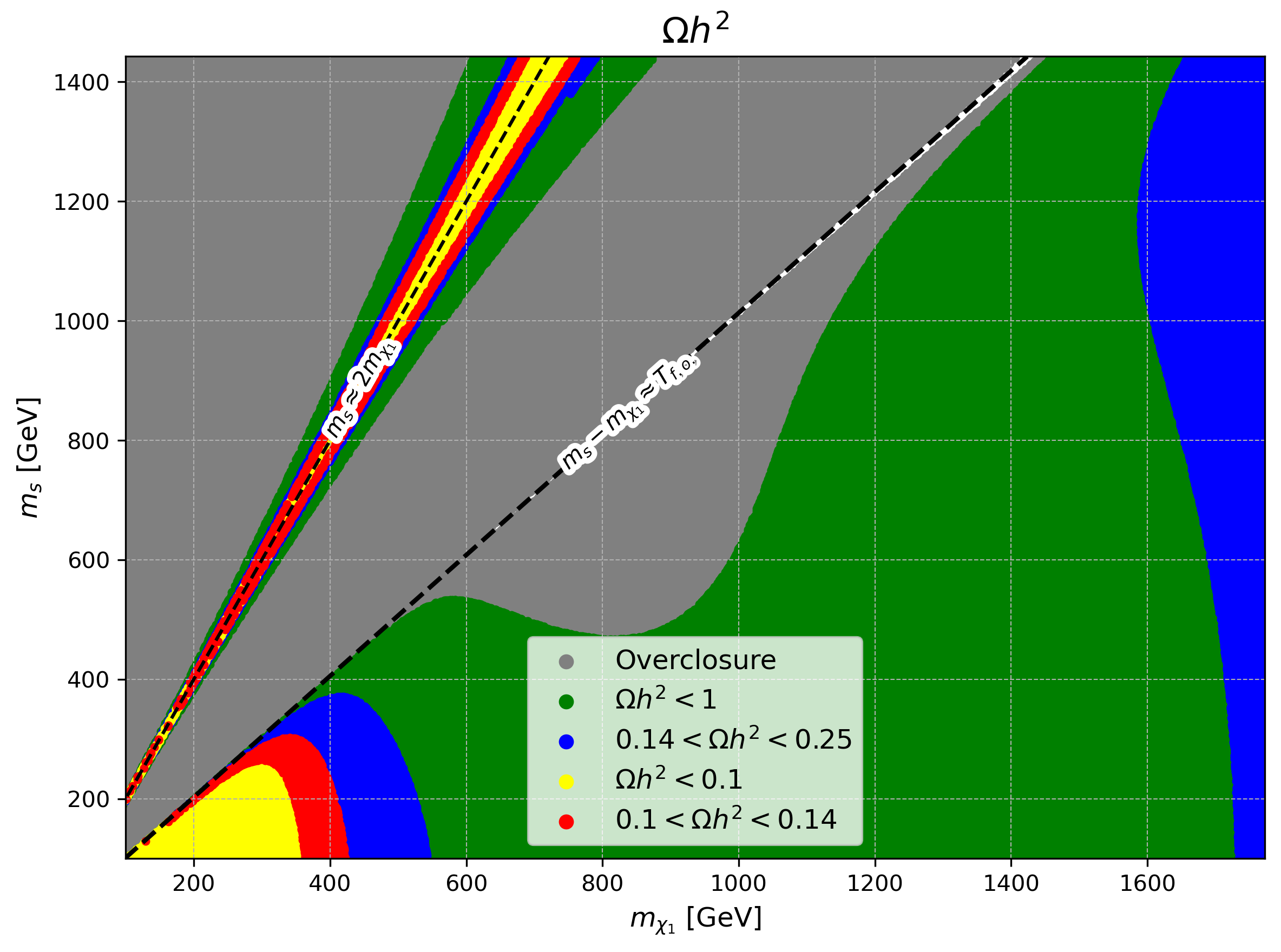}
        \caption{$v_\varphi = 0.5$ TeV, $\kappa = -6$ TeV, $\sin\theta = 0.05$}
    \end{subfigure}
    \begin{subfigure}{0.49\textwidth}
        \centering
        \includegraphics[width=\textwidth]{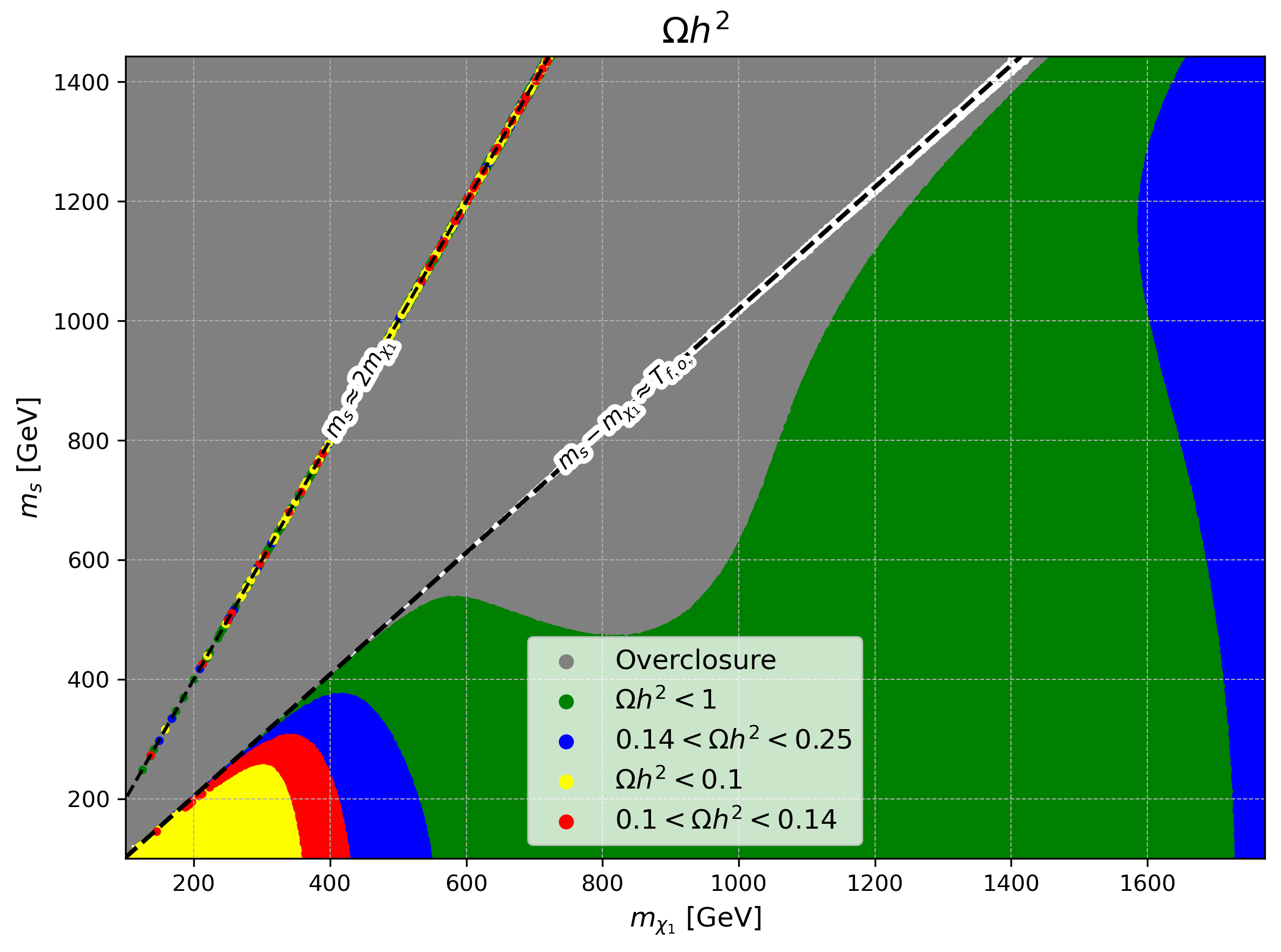}
        \caption{$v_\varphi = 0.5$ TeV, $\kappa = -6$ TeV, $\sin\theta = 10^{-3}$}\label{fig:relic_d}
    \end{subfigure}
    
    \begin{subfigure}{0.49\textwidth}
        \centering
        \includegraphics[width=\textwidth]{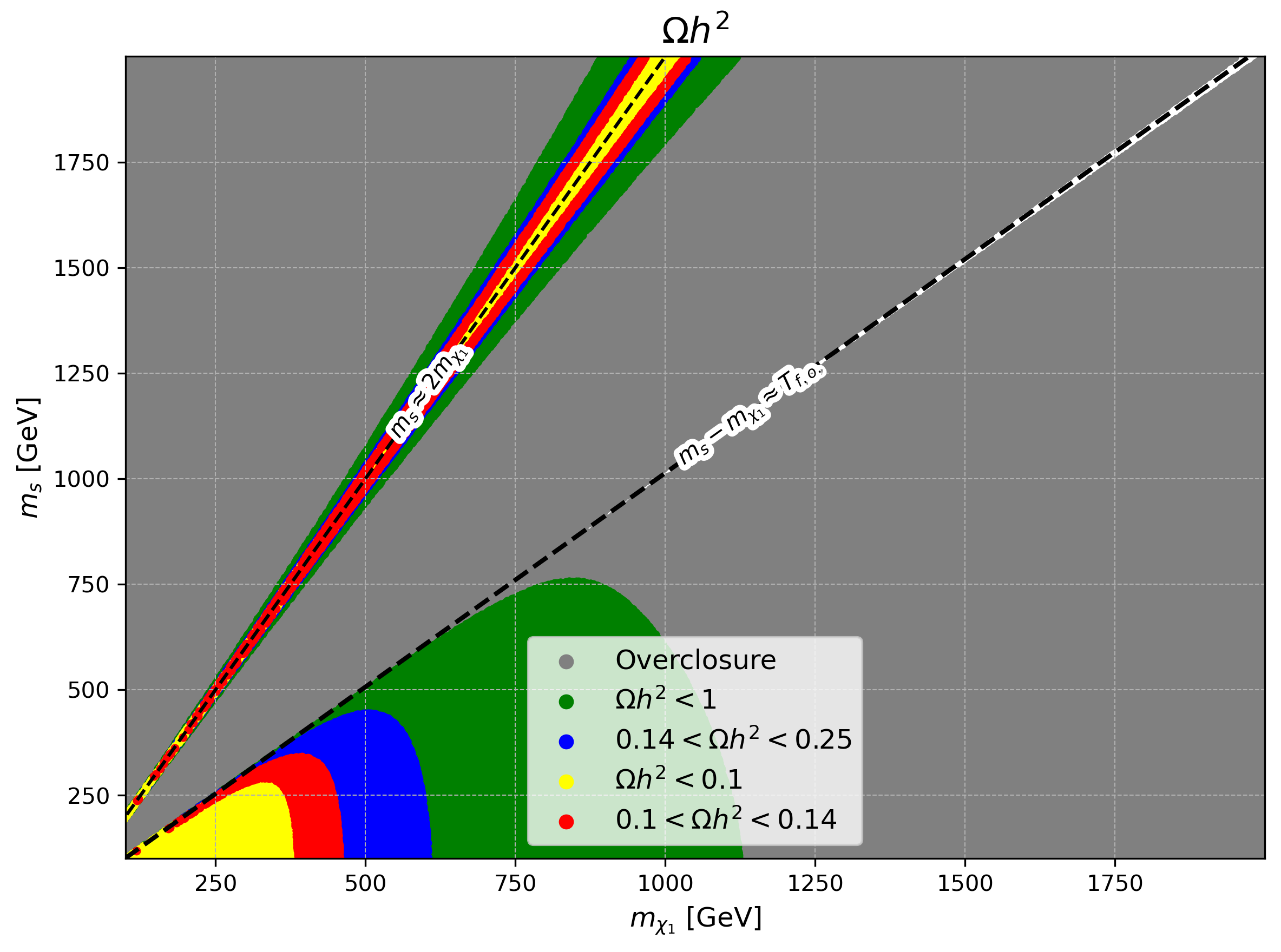}
        \caption{$v_\varphi = 1$ TeV, $\kappa = -12$ TeV, $\sin\theta = 0.05$}
        \label{fig:relic_e}
    \end{subfigure}
    \begin{subfigure}{0.49\textwidth}
        \centering
        \includegraphics[width=\textwidth]{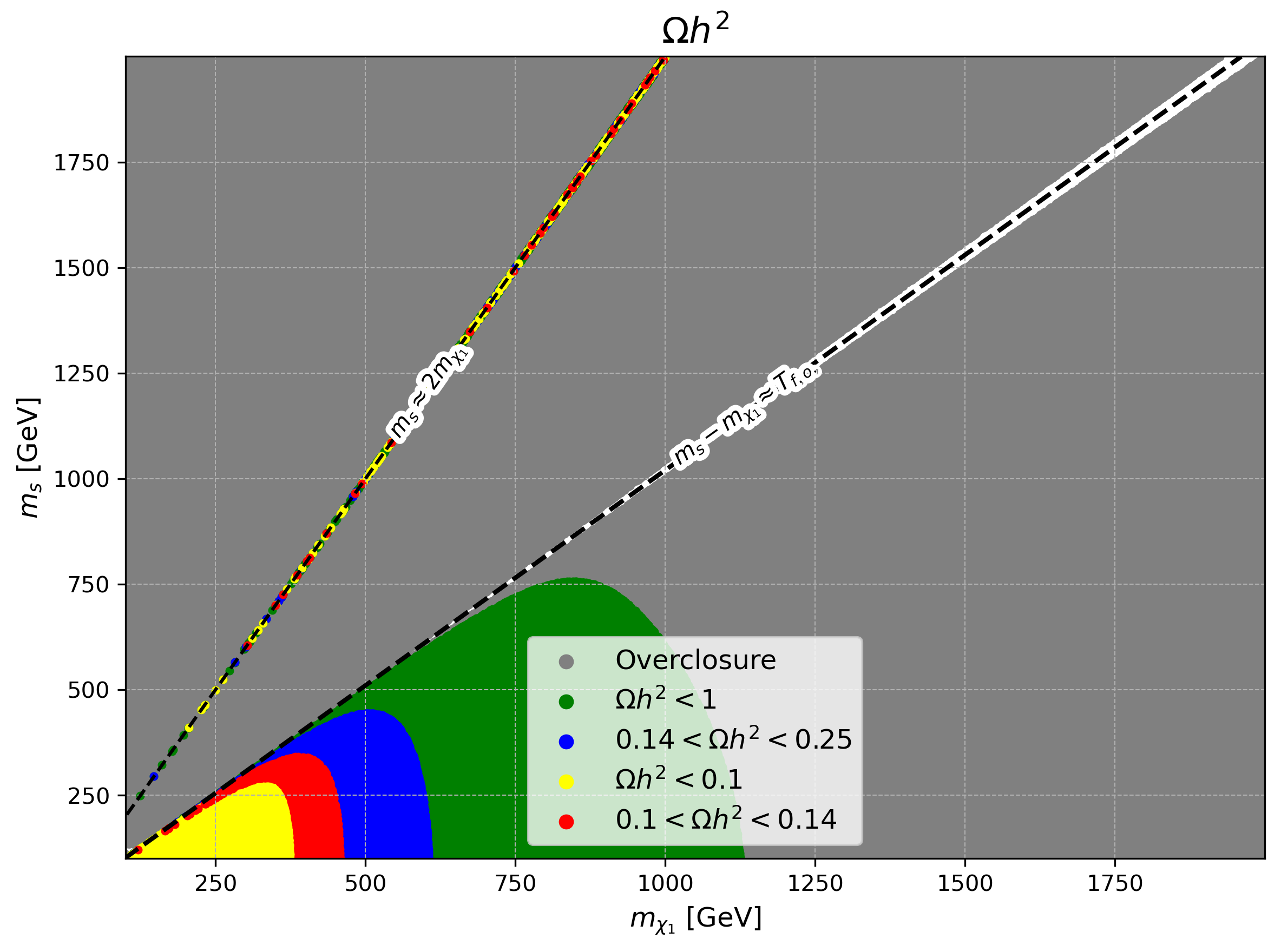}
        \caption{$v_\varphi = 1$ TeV, $\kappa = -12$ TeV, $\sin\theta = 10^{-3}$}\label{fig:relic_f}
    \end{subfigure}
    
    \caption{Relic density of dark matter particle $\chi_1$
  as a function of its mass $m_{\chi_1}$ and the scalar mass $m_s$. Different colored regions correspond to different relic density values. Each subfigure corresponds to different values of the parameters $v_\varphi, \kappa, \theta$.}
    \label{fig:relic}
\end{figure*}

Assuming all other BSM particles are heavier than the observed Higgs boson, the dominant annihilation channels and viable parameter space can be classified according to the following mass hierarchies.

\textbf{(I)~$m_{\chi_1} > m_s$, $m_{\chi_{2, 3}} \gg m_{\chi_1}$:} All final states involving $s$, $h$, SM fermions, and massive vector bosons are kinematically accessible. While all channels in Eqs.~\eqref{eq:xsec_ss}--\eqref{eq:xsec_VV} contribute, the process $\chi_1\chi_1 \to ss$ dominates when $\sin\theta \ll 1$, since it is not suppressed by factors of $\sin\theta$.

\textbf{(II)~$m_s > m_{\chi_1}$, $m_{\chi_{2,3}} \gg m_{\chi_1}$:}
In this case, the $ss$ final state is kinematically forbidden. All leading annihilation channels are suppressed by at least two powers of $\sin\theta$, and proceed via $s$-channel or $t$-channel exchange as shown in Fig.~\ref{fig:chi1_diagrams}. The dominant contributions arise from $\chi_1\chi_1 \to hs$, $hh$, $f\bar f$, and $VV$, as given in Eqs.~\eqref{eq:xsec_hh}--\eqref{eq:xsec_VV}, whenever these channels are open.

The following cases introduce important caveats to the standard thermal freeze-out picture. In these regions, the simple cross-section expressions above are no longer valid, and the relic abundance becomes highly sensitive to model parameters. We therefore provide a qualitative discussion and refer to the literature for detailed treatments.

\textbf{(III)~$m_s \gtrsim m_{\chi_1}$, $m_{\chi_{2,3}} \gg m_{\chi_1}$:} 
When $m_s$ is only slightly larger than $m_{\chi_1}$, the $\chi_1\chi_1 \to ss$ channel can still be active provided $m_s - m_{\chi_1} \lesssim T_{\rm f.o.}$. This corresponds to the \emph{forbidden dark matter} scenario~\cite{PhysRevD.43.3191,DAgnolo:2015ujb,Kopp:2016yji,Cheng:2022esn,Cheng:2023hzw}, in which annihilation is kinematically suppressed relative to the inverse
process:
\begin{equation}
\begin{split}
    &\braket{\sigma v}^{II}
        = \left[\braket{\sigma v}_{\chi_1\chi_1 \to ss} + \cdots \right] \left( \frac{n^\text{eq}_s}{n^\text{eq}_{\chi_1}} \right)^2 \\
            &\qquad \sim \left[\braket{\sigma v}_{\chi_1\chi_1 \to ss} + \cdots \right] \times  e^{-2(m_s - m_{\chi_1})/T}.
\end{split}
\end{equation}
As a result, the relic density becomes exponentially sensitive to the mass
splitting $m_s - m_{\chi_1}$. At late times, $T_0 \ll m_s - m_{\chi_1}$, which allows this scenario to evade indirect-detection constraints and leads to distinctive signatures near black holes \cite{Cheng:2022esn,Cheng:2023hzw}.

\textbf{(IV)~$m_s \simeq 2m_{\chi_1}$, $m_{\chi_{2,3}} \gg m_{\chi_1}$:} 
When the mediator mass $m_s$ lies near the threshold, annihilation proceeds dominantly through an $s$-channel resonance. In this regime, the standard velocity expansion breaks down and the mediator propagator must be treated using the Breit--Wigner form.

If $m_{\varphi} \gtrsim 2 m_{\chi_1}$, the thermally averaged annihilation cross section is enhanced in the early Universe and suppressed at late times. This can allow light dark matter to be produced thermally while evading constraints from the CMB and indirect-detection searches~\cite{Cheng:2023dau}. It can also help WIMP dark matter evade direct-detection limits~\cite{Liu:2017lpo,Arcadi:2017kky,Sheng:2023dix}. Moreover, this channel can be re-enhanced in the vicinity of black holes, with potential implications for indirect detection~\cite{Cheng:2022esn,Cheng:2023dau}.

Conversely, if $m_{\varphi} \lesssim 2 m_{\chi_1}$, the annihilation rate increases as the cosmic temperature decreases. This behavior has previously been invoked to account for the excess reported by PAMELA~\cite{Feldman:2008xs,Ibe:2008ye,Guo:2009aj}, and has also been used to explain the more recently claimed Fermi-LAT excess~\cite{Murayama:2025ihg}.

\textbf{(V)~$m_{\chi_1} \approx m_{\chi_2}$:} Coannihilation scenario~\cite{PhysRevD.43.3191}. From Eq.~\eqref{eq:massterms}, after the heavy field $\Phi$ gain vev $v_\Phi$, it can induce intra-generational mixing among $\chi_i$ if their mass spectrum is not hierarchical with $\delta m_{ij} \equiv y_{ij} \frac{v_\Phi^2}{M_\text{pl}}$. The mixing angle between interaction and mass eigenstates of heavier state $\chi_i$ and the lightest state $\chi_1$ can be parameterized as $\theta_{i1} \sim \frac{\delta m_{i1}}{m_{\chi_i} - m_{\chi_1}}$. The diagrams are \gfig{fig:chi1_coa_diagrams} with one of the external $\chi_1$ leg being replaced by a $\chi_i$ leg and a cross on the leg. Even for small Yukawa couplings $y_{ij} \ll 1$, resonant enhancement occurs when the mass splitting is comparable to the freeze-out temperature $m_{\chi_i} - m_{\chi_1} \sim \delta m_{i1} \lesssim T_\text{f.o.}$, yielding $\mathcal{O}(1)$ corrections to the annihilation rate aforementioned.

\textbf{Numerical Results:} By matching the total annihilation cross section to the value required to reproduce the observed relic abundance in Eq.~\eqref{eq:wimp_miracle}, we present the numerical results in the $m_{\chi_1}$--$m_s$ plane in \gfig{fig:relic}.

The panels are arranged such that each row corresponds to different values of $v_\varphi$ and $\kappa$, while each column corresponds to a different value of the scalar mixing angle $\sin\theta$. The upper range of the mass scan is set by the unitarity bound, whereas the lower range is chosen to be at least above half of the observed Higgs mass, $m_h/2 \simeq 62.5~\text{GeV}$. The red shaded regions indicate parameter points consistent with the observed relic density, $\Omega h^2 \simeq 0.12$. The two dashed lines correspond to $m_s - m_{\chi_1} \approx T_{\rm f.o.}$, which delineates \textbf{case III} (the forbidden dark matter scenario), and $m_s \approx 2m_{\chi_1}$, which marks \textbf{case IV} (the $s$-channel resonance annihilation scenario). These lines separate each panel into several distinct regions.

The region below the line $m_s - m_{\chi_1} \approx T_{\rm f.o.}$ corresponds to \textbf{case I} ($m_{\chi_1} > m_s$). For moderate values of $v_\varphi$, there are typically two distinct ranges of $m_{\chi_1}$ that reproduce the observed relic abundance. This feature can be understood from the dominant annihilation channel $\chi_1\chi_1 \to ss$, given in Eq.~\eqref{eq:xsec_ss}. The lower-$m_{\chi_1}$ branch is dominated by $s$-channel annihilation, while the higher-$m_{\chi_1}$ branch is dominated by $t$-channel processes. As $v_\varphi$ and $\kappa$ increase, the latter branch gradually disappears, since the $t$-channel contribution scales as $v_\varphi^{-4}$, whereas the $s$-channel contribution scales as $\kappa^2/v_\varphi^2$ for large $\kappa$.

The region above the line $m_s - m_{\chi_1} \approx T_{\rm f.o.}$ corresponds to \textbf{case II} ($m_s > m_{\chi_1}$), in which the $\chi_1\chi_1 \to ss$ channel is kinematically forbidden. In this region, annihilation proceeds dominantly through $s$-channel processes mediated by the mixed scalars, as illustrated in \gfig{fig:chi-11XX-s}. Consequently, the parameter space compatible with the observed relic abundance is localized around the sides of the resonance line $m_s \approx 2m_{\chi_1}$. This allowed region rapidly shrinks as the mixing angle $\sin\theta$ decreases, since all annihilation cross sections in this regime scale as $\sin^2\theta$. We note that in the immediate vicinity of the resonance line, both the perturbative cross-section expressions and the velocity expansion cease to be reliable. Hence, the points lying directly on the resonance line in Figs.~\ref{fig:relic_b}, \ref{fig:relic_d}, and \ref{fig:relic_f} should therefore be interpreted with caution.

\section{Implications in Direct Detection Experiments}

\begin{figure}[t]
    \centering
    \hspace*{0cm}
\includegraphics[width=8.8cm]{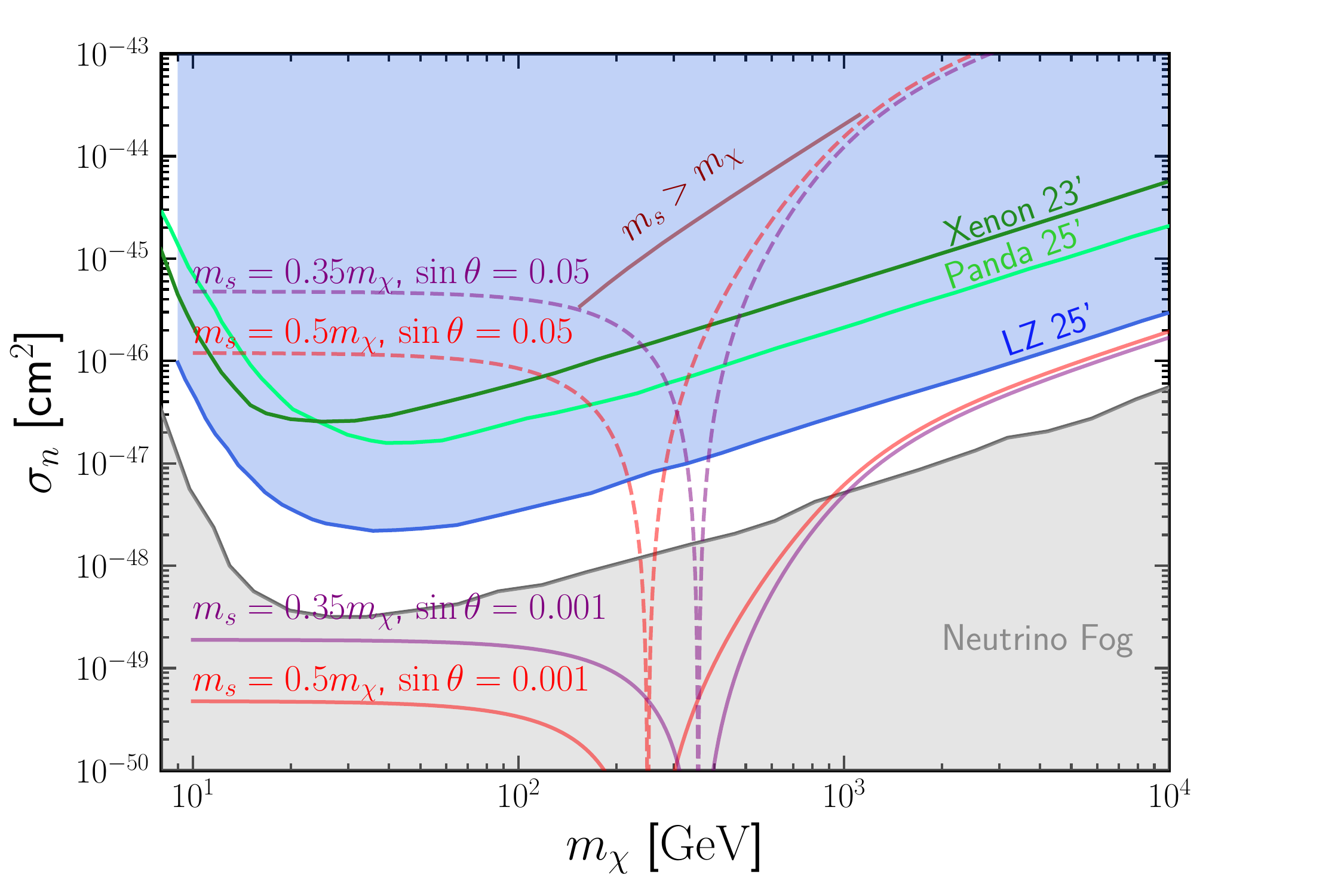}
    \caption{Comparison between the DM–nucleon scattering cross section predicted in our model and current direct-detection limits. The gray and blue shaded regions indicate the neutrino fog and the exclusions from direct-detection experiments, respectively. The red (purple) solid (dashed) curve shows the scattering cross section that reproduces the observed relic abundance $\Omega h^2 = 0.12$ for fixed ratio of $m_s = 0.5 (0.35) m_{\chi_1}$ and 
    $\sin \theta = 0.001 (0.05)$. The dark-red solid curve represents the case $m_s > m_\chi$ with 
    $\kappa = -12\,$TeV, $v_\varphi = 1\,$TeV, and $\sin \theta = 0.1$.}
    \label{DD}
\end{figure}

Via both mediators $h$ and $s$, the DM particle 
$\chi_1$ can interact with SM fermions and thus potentially leave observable signals in direct-detection experiments. For instance, the
$s$-channel annihilation process shown in \gfig{fig:chi-11XX-s} can be crossed into a scattering process, $\chi_1 + f \rightarrow \chi_1 + f$, in the $t$-channel case.
Large-scale direct-detection experiments targeting WIMP dark matter, such as PandaX~\cite{PandaX:2024qfu}, XENON~\cite{XENON:2023cxc}, and LZ~\cite{LZ:2024zvo}, primarily search for DM scattering off atomic nuclei. The underlying interaction is determined by the effective DM–nucleon coupling, i.e., $f = n(p)$, where $n(p)$ denotes the neutron (proton). After decades of continuous upgrades, these large direct-detection experiments have set very stringent bounds on the DM–nucleon scattering cross section for DM masses above $10\,$GeV, steadily approaching the so-called neutrino fog. The latest results are summarized in \gfig{DD}.

The WIMP DM in our model can likewise be probed by direct-detection experiments. Moreover, the presence of $\varphi$ endows the $\chi_1$–nucleon scattering process with distinctive features.
In the heavy-mediator approximation where the mediator mass $m_{s(h)}$ is much larger than the momentum transfer, the DM-nuclear scattering cross section is roughly,
\begin{equation}
    \sigma_{\chi n}
\simeq
    \frac{\mu_{\chi n}^2}{\pi} 
    \left[\frac{y_{\chi_1} m_n s_\theta c_\theta}{2\sqrt{2} v_\text{EW}} 
    \left( \frac{1}{m_h^2} - \frac{1}{m_s^2}\right)
    \right]^2.
\end{equation}
Here, $\mu_{\chi n} \equiv m_\chi m_n / (m_\chi + m_n)$ is the reduced mass between DM and nuclei. 
For the DM-single nuclei scattering cross section $\sigma_n$, the reduced mass is roughly the nulcei mass $\mu_{\chi n} \simeq m_n \simeq 1\,$GeV. A destructive interference between the amplitudes associated with the mediator $h$ and $s$ appears and the cross section can be greatly suppressed if $m_s \simeq m_h$. Intuitively, one can see that as the mixing angle becomes small, or as $m_s$ becomes (nearly) degenerate with the Higgs mass, our model can naturally evade the stringent direct-detection constraints~\cite{Arcadi:2016qoz}.

In realistic detectors, WIMP DM can undergo coherent scattering off the entire nucleus, with a cross section given by
\begin{equation}
    \sigma_{\chi A} \equiv 
    A^2 \frac{\mu_{\chi A}^2}{\mu_{\chi n}^2} \sigma_n,
\end{equation}
where $\mu_{\chi A} \equiv m_\chi m_A / (m_\chi + m_A)$ becomes the reduced mass of DM and the target nucleus in detector while $A$ is the corresponding atomic number. 
In Xenon-based detectors, the target atomic number is $131$ with a mass $m_A \simeq 131\,$GeV.
Both atomic number and mass enhance the DM scattering event rate a lot.
The latest bounds on the DM–nucleon scattering cross section from the three major direct-detection experiments mentioned above~\cite{XENON:2023cxc,PandaX:2024qfu,LZ:2024zvo} are shown in \gfig{DD}. These limits have been steadily approaching the so-called neutrino floor, or neutrino fog~\cite{Cirelli:2024ssz}, and are expected to reach it in the future~\cite{Wang:2023wrr,PandaX:2024muv}.

For several benchmark parameter choices, we also show in \gfig{DD} the DM–nucleon scattering cross section predicted by our model that reproduces the observed relic abundance $\Omega h^2 \simeq 0.12$, for comparison.
We first consider the case $m_s > m_{\chi_1}$. Since the annihilation channels in the early Universe are relatively involved in this regime, we directly take a representative parameter point from one of the red regions in \gfig{fig:relic_e} with ($v_\varphi = 1\,$TeV, $\kappa = -12\,$TeV, $\sin \theta = 0.1$), and convert it into the corresponding scattering cross section. The result is shown as the dark-red solid curve in \gfig{DD}. As we can see, such a case is totally excluded by direct detections. This is because the same effective 
DM-fermion four-point coupling controls both the relic production and direct detection. In this regime, essentially only a Breit–Wigner resonant-annihilation mechanism can marginally evade the direct-detection constraints~\cite{Liu:2017lpo,Arcadi:2017kky,Sheng:2023dix}.

On the other hand, the case $m_s < m_{\chi_1}$ is much richer. In this regime, the DM relic abundance is primarily controlled by the $\chi_1$-$s$ Yukawa coupling, whereas direct-detection rates are governed mainly by the $s$-$h$ mixing angle $\theta$.
In this case, the annihilation cross section for $2\chi_1 \rightarrow 2s$ is given by \geqn{eq:xsec_ss}. We assume $m_s = 0.5 m_{\chi_1}$ and 
use the relationship of $v_\varphi = m_{\chi_1} / y_{\chi_1}$, so that the annihilation cross section becomes a function of the DM mass $m_{\chi_1}$ and the Yukawa coupling $y_{\chi_1}$. By matching \geqn{eq:xsec_ss} to the thermally averaged cross section required for thermal freeze-out $\braket{\sigma v} \sim 10^{-8}\,$GeV$^{-2}$ assuming $x_{\text{f.o.}} = 25$, we determine the relation between $y_{\chi_1}$ and $m_{\chi_1}$. We then translate it into the corresponding DM–nucleon scattering cross section. The red (purple) solid (dashed) curve corresponds to the cases of $m_s = 0.5 (0.35) m_{\chi_1}$ and $\sin \theta = 0.001 (0.05)$. When the mixing angle decreases by a factor $50$, the cross section is reduced by roughly $50^2$. All curves decrease in the interference region and then rise again at higher masses. This is because obtaining the correct relic abundance for heavier DM requires a larger coupling, whereas the DM–nucleon scattering cross section depends weakly on the DM mass. Sizable portions of the parameter space along the dashed curves are already excluded for the case of $\sin \theta = 0.05$. However, due to the destructive interference between the  $h$ and $s$ mediators, the scattering cross section can become very small when $m_s \simeq m_h \simeq 125\,$GeV. For $m_{\chi_1}/m_s \sim \mathcal{O}(1)$, we therefore predict that the most plausible and currently unconstrained region corresponds to $m_{\chi_1} \sim \mathcal{O}(100)\,$GeV. However, the solid lines with $\sin \theta = 10^{-3}$ can evade almost all current direct-detection bounds. These parameter-space predictions can be further tested in future experiments.

\section{Conclusion and Discussions}

The \textit{WIMP miracle} refers to the observation that a DM particle with weak-scale mass and interactions naturally yields a thermal relic abundance close to the measured value.
This observation is phenomenologically very important, and WIMP DM remains one of the most attractive DM candidates.
Motivated by this framework, direct-detection experiments designed to search for DM scattering off nuclei have been proposed as a primary avenue to test WIMP scenarios~\cite{PhysRevD.31.3059}. To date, the leading and largest underground direct-detection experiments remain strongly optimized for WIMP searches.
In theory, SUSY has long been regarded as a well-motivated framework for WIMP~\cite{Martin:1997ns,Baer:2025srs,Baer:2025zqt,Zhang:2026eoc}. 
In particular, realizing secluded DM in supersymmetric frameworks also requires introducing additional mediators such as the Next-to-Minimal Supersymmetric Standard Model (NMSSM)~\cite{Ellwanger:2009dp}. Note that unlike the current model, the NMSSM is known to have domain wall problem if its discrete symmetry is exact~\cite{Ellwanger:2009dp}. Over the years, SUSY has been subjected to a wide range of experimental tests; however, so far it has no experimental evidence.

In this work, we propose that the intrinsic structure of the SM itself can also predict the existence of a WIMP. Owing to the distinctive structure of SM fermions, the discrete groups $\mathbb Z_4$ and $\mathbb Z_3$ are naturally embedded within the SM. We then gauge these symmetries and account for the non-perturbative effects known as the Dai–Freed anomaly, which requires the introduction of additional fermions to cancel the anomaly. Cancellation of the $\mathbb Z_4$ anomaly predicts three generation of right-handed neutrinos, while the $\mathbb Z_3$ anomaly predicts three generations of dark-sector particles $\chi_i$, and further predicts that the lightest state, $\chi_1$, is naturally a WIMP candidate. Moreover, the breaking of the discrete symmetry does not suffer from the domain wall problem.

We also present a detailed discussion of the possible thermal production mechanisms for $\chi_1$ that yield the correct relic abundance, and we map out the corresponding viable parameter space. We then compare these regions with current direct-detection constraints. We find that, because the new scalar particle 
$\varphi$ responsible for generating the $\chi$ mass is present, the DM-nucleon scattering amplitude receives contributions from the physical scalar states $s$ and $h$ that can interfere destructively. When $s$ is nearly mass-degenerate with the Higgs boson, or when the scalar mixing angle $\theta$ is small, our scenario can naturally evade existing direct-detection bounds while remaining testable in future experiments.
This naturally realizes the secluded DM scenario with a strong theoretical motivation.

The collider phenomenology of this model is also worth mentioning. As illustrated above, the freeze-out of the stable $\chi_1$ near the weak scale predicts we have a light scalar $s$ that communicates between the $\chi_1$ with the SM thermal bath, and it also mixes into the light Higgs boson. It would be interesting to see if such a scalar can be discovered soon in collider experiment such as Large Hadron Collider, whose center of mass energy can achieve $\sqrt{s} = 14$ TeV, which is enough to produce the light $s$ directly via $s$-channel resonance. The detectability depends critically on the size of the mixing angle $\theta$. 

As both the ATLAS and CMS reported a persistent anomaly in LHC's di-photon channel at 95 GeV with 1.7 $\sigma$ and 2.9 $\sigma$ significance, respectively~\cite{ATLAS:2024bjr,CMS:2024yhz}. The CMS also found a 95 GeV anomaly with 2.6 $\sigma$ significance in $\tau\tau$ channel~\cite{CMS:2022goy}. This anomaly coincides with the allowed and preferred range of the light scalar $s$ in our model. We stress that from \gfig{DD}, a 95 GeV $s$ can situate very well within the interference region that could simultaneously fulfill the observed relic abundance while allowed by the direct detection limit with a mixing angle that can even be as large as $\sin\theta \sim 0.01$. We leave a detailed collider phenomenology study on the potential of using our model to address the 95 GeV LHC anomaly while explaining the relic abundance in an upcoming future work .

\section*{Acknowledgements}

We thank Kazuya Yonekura for the useful discussion on the Dai-Freed anomalies. J. S. and T. T. Y. thanks Yu Cheng for the discussion in the early stage of this work. J. S. is supported by the Japan Society for the Promotion of Science (JSPS) as a part of the JSPS Postdoctoral Program (Standard) with grant number: P25018. T. T. Y. is supported by the Natural Science Foundation of China (NSFC) under Grant No. 12175134, MEXT KAKENHI Grants No. 24H02244.
Both J. S. and T. T. Y are supported
by the World Premier International Research Center Initiative (WPI), MEXT, Japan (Kavli IPMU). K. Z. gratefully acknowledges support from the Avenir Foundation.

\begin{appendix}
\section{Appendix A -- Origin of the $\mathbb Z_4 \times \mathbb Z_3$ discrete symmetry}\label{appendix-discrete}

We explain in more detail how the discrete symmetry
\begin{equation}
    \mathbb Z_{12} \simeq \mathbb Z_4 \times \mathbb Z_3
\end{equation}
arises in our setup and why it is preserved even by the non-perturbative effects of the SM gauge interactions.

Consider the $SU(3)_c$ instanton. For one family, there are four fermion zero modes, $u$, $d$, $\bar u$, $\bar d.$ In our model, these four fields all have the same $\mathbb Z_4$ charge from \gtab{fermion},
\begin{equation}
    Q_{\mathbb Z_4}(u)
        =Q_{\mathbb Z_4}(d)
        =Q_{\mathbb Z_4}(\bar u)
        =Q_{\mathbb Z_4}(\bar d)
        =1 \quad (\mathrm{mod}\ 4).
\end{equation}
Therefore, the total $\mathbb Z_4$ charge of the instanton vertex is
\begin{equation}
    1 + 1 + 1 + 1 
        = 4 
        \equiv 0 \qquad (\mathrm{mod}\ 4).
\end{equation}
Hence the $SU(3)_c$ instanton preserves a $\mathbb Z_4$ symmetry.

Next, consider the $SU(2)_L$ instanton. For one family, there are again four fermion zero modes: three quark doublets $q$ and one lepton doublet, $\ell$. In our charge assignment from \gtab{fermion},
\begin{equation}
    Q_{\mathbb Z_4}(q)
        =Q_{\mathbb Z_4}(\ell)
        =1 \qquad (\mathrm{mod}\ 4),
\end{equation}
and therefore the total charge of the $SU(2)_L$ instanton vertex is again
\begin{equation}
    1+1+1+1 
        = 4 
        \equiv 0 \qquad (\mathrm{mod}\ 4).
\end{equation}
Thus the same $\mathbb Z_4$ symmetry is preserved by the $SU(2)_L$ non-perturbative effect. 
As a result, both $SU(3)^2$-$\mathbb Z_4$ and $SU(2)^2$-$\mathbb Z_4$ mixed gauge anomalies disappear.
Before the recognition of the Dai–Freed anomaly, pure anomalies of discrete symmetries were generally not analyzed, since there are no corresponding continuous gauge fields. When one considers the mixed gravity$^2$-$\mathbb Z_4$ anomaly, one finds that a single generation contains 15 Weyl fermions, as shown by the $T(10)$ and $\bar F(5^*)$ representations. Therefore, right handed neutrino should be introduced to cancel this anomaly.

The $\mathbb Z_3$ factor simply follows from the existence of three families. As illustrated above, in both the cases of $SU(3)_c$ and $SU(2)_L$, since there are four zero modes for one family, the total number of zero modes is $12$ for three families. If each chiral fermion carries unit $\mathbb Z_3$ charge, then the instanton vertex has total charge
\begin{equation}
    12 \equiv 0 \qquad (\mathrm{mod}\ 3).
\end{equation}
Hence the non-perturbative effects also preserve a $\mathbb Z_3$ symmetry.

In conclusion, the instanton effects preserve both $\mathbb Z_4$ and $\mathbb Z_3$, and we obtain
\begin{equation}
    \mathbb Z_{12} = \mathbb Z_4 \times \mathbb Z_3
\end{equation}
as an exact discrete symmetry even at the non-perturbative level.

Because this symmetry is preserved by both perturbative interactions and the SM instanton effects, it is natural to regard it as a discrete gauge symmetry. In our framework, this is highly non-trivial. Gauging $\mathbb Z_4$ requires cancellation of the corresponding Dai--Freed anomaly, which is achieved by introducing three RHNs $N_i$. 
Likewise, anomaly cancellation for $\mathbb Z_3$ requires three chiral fermions $\chi_i$, which are precisely the dark sector fermions, naturally serving as WIMPs.

\end{appendix}

\bibliographystyle{utphys}
\bibliography{ref}

\vspace{15mm}
\end{document}